\documentstyle[11pt]{article}
\begin{document}
\title{Half Quantization}
\author{Nuno Costa Dias \\ {\it Grupo de Astrof\'{\i}sica e Cosmologia} \\ {\it Departamento de F\'{\i}sica - Universidade da Beira Interior}\\ {\it 6200 Covilh\~{a}, Portugal}}
\date{}
\maketitle

\begin{abstract}
A general dynamical system composed by two coupled sectors is considered.
The initial time configuration of one of these sectors is described by a set of classical data while the other is described by standard quantum data. These dynamical systems will be named half quantum. 
The aim of this paper is to derive the dynamical evolution of a general half quantum system from its full quantum formulation. 
The standard approach would be to use quantum mechanics to make predictions for the time evolution of the half quantum initial data.
The main problem is how can quantum mechanics be applied to 
a dynamical system whose initial time configuration is not 
described by a set of fully quantum data.
A solution to this problem is presented and used, as a guideline
to obtain a general formulation of coupled classical-quantum dynamics.
Finally, a quantization prescription mapping 
a given classical theory to the correspondent half quantum one
is presented.
\end{abstract}

\section{Introduction}

Quasiclassical dynamics \cite{anderson1}, hybrid dynamics \cite{diosi1} and, in this paper, half quantum mechanics are some of the several attempts \cite{aleksandrov,boucher,halliwell1,halliwell2} to obtain a consistent formulation of coupled classical-quantum dynamics.
The motivation to develop such a theory comes from a variety of different sources. The theory is expected to make important contributions to clarify the measurement procedure in quantum mechanics, where one would like to obtain an analytic description of the wave function collapse \cite{nature,sherry,diosi2}. Closely related is the problem of developing a consistent quantization 
procedure for closed dynamical systems \cite{hartle1,halliwell3}. Other important applications are expected. These include semiclassical gravity, quantum field theory in curved space time and quantum cosmology \cite{halliwell1,nature,hartle1,wald}.

Two main approaches to the problem have been followed: In \cite{anderson1,aleksandrov,boucher} a set of axioms defining the 
quasiclassical dynamics were proposed and motivated in terms of the consistency of the thus resulting theory. On the other hand there is the deductive approach where the intention is to derive the classical-quantum dynamics from quantum mechanics \cite{diosi1,halliwell1,halliwell2}.

In this paper we shall follow this second approach. We assume that, just like classical mechanics, half quantum mechanics is an approximate description of quantum mechanics that derives its validity from reproducing, "in some appropriated sense", the predictions of the underlying theory of quantum mechanics. 

Our approach will be as follows: A general half quantum system is composed by two coupled sectors. One of these sectors is named classical and the other quantum. The initial data for an half quantum system is given by a set of classical data $O_i(t=0)$ for the classical sector plus a standard quantum 
data, say an initial time wave function $|\phi^Q>$ for the quantum sector.
The important issue is how can quantum mechanics be applied to a dynamical system whose initial time configuration is not described by a set of fully quantum data. 
To solve this problem we will convert the half quantum initial data into a 
fully quantum one. More precisely, we will determine a class of wave functions $|\phi>$ which are consistent with the half quantum initial data. We will be able to do this by using a classicality criterion that was presented in a related paper \cite{nuno1} and which proved to work out successfully when the intention was to study the consistency between the full classical and full quantum descriptions of a general dynamical system.
We can then use quantum mechanics to obtain the time evolution of this class 
of quantum initial data. The predictions of quantum mechanics, i.e. the time evolution of the class of wave functions $|\phi>$, will not be completely determined. This is so because we do not have a single initial data wave function, but instead we are calculating the time evolution for a class of initial data wave functions. Therefore, quantum mechanics provides a set of predictions inside an error interval.

The main result is then that these predictions might be fully recovered by an appropriated formulation of classical-quantum dynamics, which will be named 
half quantum mechanics.    
In this formulation the dynamical system is not fully quantized, the classical data describing the initial time configuration of the classical sector is explicitly used
and dynamics is obtained as the time evolution of the classical and quantum initial data. 
Still, we are able to recover the predictions of quantum mechanics for the time evolution of the class of wave functions $|\phi>$. This is the desired result. It means that the half quantum framework is derived as the appropriated limit 
of quantum mechanics.

We will find that the theory derived here is just the same one that was postulated by Boucher and Traschen in \cite{boucher}. The approach however, is rather different. In that paper the theory was motivated in terms of the properties one would like to see satisfied by a theory of coupled classical-quantum dynamics.

Our derivation presents some interesting features: i) it explicitly provides the degree of precision of the half quantum predictions i.e. it tells us about the degree of consistency between the half quantum and the full quantum predictions. ii) It states what type of initial data and dynamical behaviour a system should have so that it can be described by the half quantum framework. iii) It settles down a general procedure, with assumptions kept to a minimum, to develop other, eventually more consistent or better-behaved, classical-quantum dynamics frameworks. iv) It provides a half quantization procedure mapping the classical formulation of a given dynamical system to its half quantum formulation.

\section{From quantum mechanics to half quantum mechanics}

Let us settle down the preliminaries: we are given a dynamical system with 
$N+M$ degrees of freedom. 
The $N$ represents the number of degrees of freedom of
the quantum sector, while the $M$ concerns the
classical sector. The phase space of the classical formulation of the system 
is spanned by a set of canonical variables $\{q_k,p_k\}, k=1..(M+N)$ or more succinctly just designated by $O_k, k=1..2(M+N)$. The classical sector canonical variables are denoted by $(q_i,p_i), i=1..M$ or just by $O_i, i=1..2M$ and the quantum sector canonical variables by $(q_{\alpha},p_{\alpha}), \alpha = (M+1)..(M+N)$ or $O_{\alpha}, \alpha =(2M+1)..2(M+N)$. 
The total phase space is assumed to have a structure given by $T^{\ast}M_1 \otimes T^{\ast}M_2$ where $T^{\ast}M_1$ is the classical sector phase space 
and $T^{\ast}M_2$ is the quantum sector phase space. 

By performing the Dirac quantization \cite{dirac1,dirac2} we obtain the quantum formulation of the dynamical system. We also supply a complete set of commuting observables (CSCO).
We will take the CSCO to be 
$\{\hat{q}_{i},\hat{q}_{\alpha}\}$.
The set of common eigenvectors of the CSCO spans the Hilbert space ${\cal H}
={\cal H}_1 \otimes {\cal H}_2$.
Taking into account the structure of the Hilbert space the general  
 eigenvector might be written as $
|k_1,...k_N>|z_1,...z_M> $ where the $k$s are eigenvalues of the operators 
$\hat{q}_{\alpha}$ and the $z$s are those of the operators $\hat{q}_i$.

The goal is now to use the full quantum formulation of the dynamical system to study the time evolution of the half quantum initial data. 
This is far from being straightforward, the problems one encounters being
closely related to those that emerge when one wants to study the dynamics of a classical 
system using the framework of quantum mechanics \cite{nuno1}. As in that
case, the first problem is how to use the 
half quantum initial data to produce fully quantum initial data for the quantum theory.
This problem shall be approached in this section.

\subsection{From quantum mechanics to half quantum mechanics - Kinematics}

The half quantum dynamical system is composed by two sectors. The initial 
time configuration of one of these sectors is described by a set of classical data.
That is a value $O^0_i$ and an error margin $\delta_i$ is assigned to each classical sector observable $O_i$. The aim is to convert these classical data 
into a full quantum one, $|\phi^c> \in {\cal H}_1$. Clearly, not all wave functions $|\phi^c>$ will be suitable. We are looking for a class of wave functions $|\phi^c> \in {\cal H}_1$ providing a description of the initial time configuration consistent with the classical description $(O^0_i,\delta_i)$. To obtain this class of wave functions we impose that $|\phi^c>$ should satisfy a set of classicality conditions that was defined and studied in \cite{nuno1}. More precisely, we require $|\phi^c>$ to be $L$-order classical ($L\in {\cal N}$) with respect to the classical data $(O^0_i,\delta_i)$. The higher the order of classicality $L$, the bigger will be the degree of consistency between the classical and the quantum descriptions.
We will not fix the value of $L$. In fact, 
$L$ is to be one of the parameters of the formalism and latter we will find 
that its value is related to the precision of the half quantum predictions.

Let us make a brief review of the definition of the classicality criterion. Let $O_k(t)$ be the classical time evolution of an arbitrary fundamental 
observable (belonging to the classical or to the quantum sector)
and let $S_{ia}$ be {\it any} sequence of classical sector observables 
$S_{ia}=O_{i1},O_{i2}....O_{in}$ - associated to a sequence  
$1 \le i_a \le 2M, a=1..n$ ($n$ is arbitrary) -
such that:
\begin{equation}
\frac{\partial^n O_k(t)}{\partial S_{ia}}=
\frac{\partial^n O_k(t)}{\partial O_{i1} ....\partial O_{in}} \not=0  ,
\end{equation}
for some $k=1..2(N+M)$. With all sequences satisfying the former relation 
we can obtain a set of 
mixed error kets (the reader should refer to the appendix for 
the relevant definitions):
\begin{equation}
|E_{S_{ia}}>=(\hat{O}_{i1}-O^0_{i1})(\hat{O}_{i2}-O^0_{i2})....
(\hat{O}_{in}-O^0_{in})|\phi^c>,
\end{equation}
where the quantities $O^0_{ia}$ refer to the values of the
corresponding observables $O_{ia}$ at the initial time.
The classical sector initial time wave function $|\phi^c>$ will be {\it 1st-order classical} 
if it satisfies:
\begin{equation}
<E_{S_{ia}}|E_{S_{ia}}> \le (\delta_{S_{ia}})^2 = \delta_{{i1}}^2  
\delta_{{i2}}^2....\delta_{{in}}^2 , 
\end{equation}
for all the sequences $S_{ia}$ determined in (1). In the former equation 
$\delta_{{ia}}$ are the error margins associated to the classical 
initial data.  
Notice that given the classical initial data and its error margins the former inequalities 
constitute a set of requirements on the functional form of the wave 
function $|\phi^c>$. To go further we consider the $L$-order sequences $S^L_{ia}=S_{ia}S_{i^{\prime}a}...S_{i^{\prime \prime}a}$ constituted by $L$ arbitrary 1st-order sequences $S_{ia}$ (determined in (1)) and write the system of inequalities (3) for these sequences. If the wave function $|\phi^c>$ satisfies (3) for all possible $L$-order sequences then we say that $|\phi^c>$ is $L$-order classical. 

The set of all $L$-order classical wave functions is the class of wave functions that we wanted to determine. It is worth noticing (appendix: result a)) that all $L$-order classical wave functions $|\phi^c>$ satisfy the following property: in the representation
of any of the observables $\hat{O}_i$, they have a probability of at least $p$ 
confined to the interval $I_i=[O^0_i-\delta_{i}/(1-p)^{1/2L},  
O^0_i+\delta_{i}/(1-p)^{1/2L}]$, that is $ \sum_{a_i \in I_i ,n}
|<a_i^n|\phi^c>|^2 \ge p$ for all $i=1..2M$ and where $|a^n_i>$ is a general 
eigenvector of the operator $\hat{O}_i$ with associated eigenvalues
$a_i$ and degeneracy index $n$. By simple inspection of the former result we notice that the higher the degree of classicality $L$, the bigger is the confinement of the probabilistic distribution function associated to $|\phi^c>$, around the classical intervals $[O^0_i-\delta_i,O^0_i+\delta_i]$.  

Given the classical initial data, the degree of classicality is a statement about the quantum mechanical description of a given configuration of the dynamical system. It tell us that the wave function $|\phi^c>$ satisfies some properties. In the context of this paper it may be worth thinking about the classicality criterion in an equivalent but slightly different perspective: the degree of classicality can be seen as the degree of {\it validity} of the classical description of a given configuration of the dynamical system. The classical description is {\it valid} up to some degree $L$ if the true, physical configuration of the dynamical system is given by a wave function $|\phi^c>$,  $L$-order classical with respect  
to that classical description.

In conclusion: the true physical configuration of the classical sector is given by a wave function $|\phi^c>$. However, we are only given the classical imprecise description $(O_i^0,\delta_i)$ and thus we should not assume that we know $|\phi^c>$ completely. 
If we assume that the classical initial data is $L$-order valid then any wave function belonging to the class of $L$-order classical wave functions can be, up to what we know, the true physical configuration of the system. Therefore, the classical sector initial configuration is properly described, not by a single wave function, but by the class of $L$-order classical wave functions.  

The initial time configuration of the other sector of the half quantum system is described by standard quantum data. That is we supply a completely fixed initial time wave function:
\begin{equation}
|\phi^Q>=\sum_{k_1,...k_N} C_{k_1,...k_N} |k_1,...k_N> .
\end{equation} 
The final step is to put the two sectors together and obtain
the total wave function. 
To make it simple we assume that there is no kinematical coupling between the two sectors and thus the initial data wave function is of the form:
\begin{equation}
|\phi>=|\phi^Q>|\phi^c>=
\sum_{k_1,...k_N} C_{k_1,...k_N} |k_1,...k_N>|\phi^c>,
\end{equation} 
where $|\phi^c>$ is a $L$-order classical wave function.

\subsection{From quantum mechanics to half quantum mechanics - Dynamics}

The goal now is to obtain the time evolution of the initial data wave 
function (5). To do this let us work in the Heisemberg picture and let us
calculate the full quantum time evolution of an arbitrary fundamental observable  
$\hat{O}_k$:
\begin{equation}
\hat{O}_k(t)=\sum_{n=0}^{\infty} \frac{1}{n!} 
\left( \frac{t}{i\hbar} \right)^n
[...[\hat{O}_k,\hat{H}]...,\hat{H}].
\end{equation} 
Let us designate the 
general operator $\hat{O}_k(t_0)$ just by $\hat{A}$. The aim is then to study 
the functional form of the initial data wave function in the representation of 
$\hat{A}$. The first step is to write 
the general observable $\hat{A}$ as a sum of multiple
products of the fundamental observables:  
\begin{equation}
\hat{A}=\sum_j \hat{A}^c_j \hat{A}^Q_j \quad:
\left\{ \begin{array}{ccc}
\hat{A}^c_j=c_j \prod_{a=1}^{n_j}\hat{O}^{c}_{i_a(j)}\\
\hat{A}^Q_j=\prod_{b=1}^{m_j}\hat{O}^{Q}_{\alpha_b(j)}
\end{array} \right.
\end{equation}
where for each $j$ the sets of 
coefficients $i_a(j)$ and $\alpha_b(j)$ are two 
sequences, the first one 
having values in $\{1..2M\}$ and the second one in $\{2M+1...2(M+N)\}$
and $c_j$ are complex parameters that may depend on time.
Let us proceed naively and try to obtain predictions for the outputs
of a measurement of $\hat{A}$. Let then $|a_i^n>$ be the general eigenvector
of $\hat{A}$ with associated eigenvalue $a_i$ and degeneracy index $n$.
Using the standard prescription the predictions are given by the 
set of pairs $(a_i,P(a_i))$ where $P(a_i)$ is the probability of 
obtaining the value $a_i$ from a measurement of $\hat{A}$, i.e.
$P(a_i)=\sum_n |<a_i^n|\phi>|^2$. 
We easily realise that we have a problem. In fact we do not know
$|\phi>$ completely and so the calculation of $P(a_i)$ is, 
to say the least, not straightforward. 

To circumvent the problem we introduce a new operator $\hat{B}$ obtained by applying a map $V_0$, named unquantization, to the operator $\hat{A}$. This map $V_0$ is defined as a trivial extension of the full unquantization map (mapping quantum operators to full classical observables) that was defined and studied in a related paper. Let us then present the definition of $V_0$:

\bigskip

\underline{{\bf Definition I}} {\bf - First unquantization map} \\
Let ${\cal A}({\cal H})$ be the algebra of linear operators 
acting on the Hilbert space ${\cal H}={\cal H}_1 \otimes {\cal H}_2$ and let ${\cal S}$ be the algebra of $C^{\infty}$ functionals ${\cal S}=\{ f:T^{\ast}M_1 \longrightarrow 
{\cal A}({\cal H}_2) \}$. The unquantization $V_0$ is a map from ${\cal A}({\cal H})$ to ${\cal S}$ that satisfies the following rules (we use the notation of (7)):\\
{\bf 1)} $V_0(\sum_j \hat{A}^c_j \hat{A}_j^Q)=\sum_j V_0(\hat{A}^c_j)V_0(\hat{A}^Q_j)$\\
{\bf 2)} $V_0(\hat{A}^Q_j)=\hat{A}^Q_j$\\
{\bf 3)} $V_0(\hat{A}^c_j)=A^c_j$. The unquantization map that take us from $\hat{A}_i^c$ to $A_i^c$ was defined in \cite{nuno1} when the intention was to derive the full classical observable that corresponds to a general quantum operator. The following steps defined this procedure: 
i) $\hat{A}^c_j$ should be expanded as a sum of a hermitian operator and an anti-hermitian one, 
ii) all antisymmetric terms of $\hat{A}^c_j$ should then be executed i.e. all the commutators present in $\hat{A}^c_j$ should be calculated, 
iii) finally, given $\hat{A}^c_j$ displayed in an order satisfying the two previous requirements we perform the substitution of the quantum fundamental operators present in $\hat{A}^c_j$ by the corresponding classical canonical variables, i.e. if $\hat{A}^c_j=F(\hat{O}_i)$ where $F$ satisfies the order requirements i) and ii) then $A^c_j=F(O_i)$.

By applying $V_0$ to $\hat{A}$ we get:
\begin{equation}
\hat{B}=\sum_j A^c_j \hat{A}^Q_j .
\end{equation}
Notice that $V_0(\hat B )$ is not completely well defined ($V_0$ is not univocous). In fact there are several different orders in which we can display $\hat{A}$ all of them satisfying the requirements i) and ii) but producing different operators $\hat{B}$. This ambiguity will be discussed in detail in the next section. However we should point out that all future results of this section are valid for all operators $\hat{B}$ obtained from unquantizing the same operator $\hat{A}$.

The aim now is to use a representation induced by $\hat{B}$ to
obtain some knowledge about the properties of the initial
data wave function $|\phi>$ in the representation of $\hat{A}$.
To do this some preliminary work is needed.

\subsubsection{Representations induced by $\hat{B}$}

Let us consider the general state $|\psi>=|\phi^c>|\psi^Q>$, where $|\phi^c>$ is the classical sector initial data wave function, that we assume to be $L$-order classical with respect to the classical initial data $(O^0_i,\delta_i)$, and $|\psi^Q>$ is an arbitrary quantum sector wave function. 
The first step will be to obtain the value of $|<\psi|(\hat A -\hat B)^{2L}|\psi>|^{1/2L}$ as a function of the half quantum initial data and of the half quantum operator. The relevance of this result will become clear latter on.\\
{\bf a)} Explicit form of $|<\psi|(\hat A -\hat B)^{2L}|\psi>|^{1/2L}$:\\
Let $A_j^c=V_0(\hat A^c_j)$. The following relation is valid up to a correction term of the order of $\hbar^2$ \cite{nuno1}:
\begin{equation}
\hat{A}^c_j-A^c_j=
\sum_{i=1}^{2M} 
\frac{\partial A^c_j}{\partial O_i}(\hat{O}_i-O_i)
+\frac{1}{2}\sum_{i,k=1}^{2M}\frac{\partial^2 A^c_j}
{\partial O_i \partial O_k}
(\hat{O}_i-O_i)(\hat{O}_k-O_k) +...
\end{equation}
Using the former equation we have:
\begin{eqnarray}
 (\hat{A} &-& \hat{B})^L  =  \left\{\sum_j(\hat{A}^c_j-A^c_j)\hat{A}^Q_j \right\}^L \nonumber\\
& = & \left\{ \sum_{i=1}^{2M} 
\frac{\partial \hat B}{\partial O_i}(\hat{O}_i-O_i)
+\frac{1}{2}\sum_{i,k=1}^{2M}\frac{\partial^2 \hat B}
{\partial O_i \partial O_k}
(\hat{O}_i-O_i)(\hat{O}_k-O_k) +...+\sum_j c_j\hbar^2 \hat{\epsilon}_j \hat{A}^Q_j  \right\}^L ,
\end{eqnarray}
where we explicitly included the correction term $c_j\hbar^2 \hat{\epsilon}_j$. 
Up to the lowest order we get:
\begin{equation}
(\hat{A}-\hat{B})^L  = \sum_{i_1=1}^{2M}...\sum_{i_L=1}^{2M} \prod_{s =1}^L 
\frac{\partial \hat B}{\partial O_{i_s}}(\hat{O}_{i_s}-O_{i_s})
+...+{\cal O}(\hbar)^2 .
\end{equation}
The relation (9) was derived and discussed in detail in \cite{nuno1}. There we point out that (9) is exactly valid only if $A^c_j$ is obtained from a total symmetric form of $\hat{A}^c_j$. This is not the general case if we use the unquantization map $V_0$ to obtain 
$A^c_j$. However we also saw in \cite{nuno1} that if we use the map $V_0$ - in which case $A_j^c$ is obtained from $\hat{A}^c_j$ displayed in an order that does not contain  antisymmetric components (see definition I) - then the difference between the two sides of eq.(9) is
given by $c_j\hbar^2 \hat{\epsilon}_j$, where $c_j$ is the numerical factor in $\hat{A}_j^c$  (7) and $\hat{\epsilon}_j$ is the "operator error" proportional to a sum of products of monomials $(\hat{O}_i -O_i)$, each of the products having at the most $n_j-2$ terms (check eq.(7) for the meaning of $n_j$). This error term was explicitly included in the expansion (10). Typically, the contribution of the term proportional to $c_j\hbar^2$ is meaningless when compared to the terms proportional to the derivatives of $\hat{B}$. However in some artificial examples this may not be the case. Consider for instance: $\hat{A}=(\hat{x}^c \hat{y}^c \hat{z}^c +\hat{z}^c \hat{y}^c \hat{x}^c)
\hat{A}^Q -(\hat{y}^c \hat{x}^c \hat{z}^c +\hat{z}^c \hat{x}^c \hat{y}^c)\hat{A}^Q $, 
where $\hat{x}^c, \hat{y}^c, \hat{z}^c$ are hermitian, classical sector operators of an arbitrary system.
We have $V_0(\hat{A})=xyz \hat{A}^Q- xyz \hat{A}^Q=0 = \hat{B}$ and therefore, in this case, $\hat{A}-\hat B=\sum_jc_j\hbar^2\hat{\epsilon}_j\hat{A}^Q_j$ which, in general, is not zero. 
The problem lies, of course, in the order in which $\hat{A}$ is displayed before we apply the map $V_0$. One should impose the restriction that $\hat{A}$ can not be displayed in an order in which some unresolved commutatores are present. One easy way to check that this is the case is precisely by  
comparing the magnitude of $\hat{B}$ (the numerical factors in $\hat{B}$) with the magnitude of $c_j\hbar^2$, where the $c_j$ are the numerical factors of $\hat{A}$. For physical relevant examples (physical Hamiltonians and observables), and namely for the time evolution of a general quantum observable, it is easy to verify if the original operator $\hat{A}$ is in an adequate order (this is in fact the typical case), and thus upon unquatization, one has {\it magnitude}$(\hat{B}) \propto c_j \hbar^0 >> c_j\hbar^2$. Therefore, and keeping the caution remark in mind, we shall take the result (11) to be exactly valid.  

To proceed we apply the expansion (11) to the state $|\psi>$.  
To the first order we have:
\begin{equation}
(\hat A -\hat B)^{L}|\psi>  = \sum_{i_1=1}^{2M} ... \sum_{i_L=1}^{2M}
|E(\hat{O}_{i_1},..,\hat{O}_{i_L},\phi^c,O_{i_1},..,O_{i_L})>  \prod_{s=1}^L
\frac{\partial \hat{B}}{\partial O_{i_s}} |\psi^Q>  +...
\end{equation}
Using (12) just up to the lowest order we get:
\begin{eqnarray}
\lefteqn{ |<\psi|(\hat A -\hat B)^{2L}|\psi>|  \le }\\
 \sum_{i_1=1}^{2M} .. \sum_{i_L=1}^{2M} &&
\sum_{k_1=1}^{2M} .. \sum_{k_L=1}^{2M} 
|<E_{O_{k1}..O_{kL}} |E_{O_{i1}..O_{iL}}>| 
 |<\psi^Q|\frac{\partial \hat{B}^{\dagger}}{\partial O_{k_1}}..
\frac{\partial \hat{B}^{\dagger}}{\partial O_{k_L}} 
\frac{\partial \hat{B}}{\partial O_{i_1}}.. 
\frac{\partial \hat{B}}{\partial O_{i_L}}|\psi^Q>|+...   \nonumber
\end{eqnarray}
and using the Shwartz inequality $L$ times, the relation (3) and disregarding the contributions of terms proportional to $\hbar^2$ or smaller we get:
\begin{equation}
|<\psi|(\hat A -\hat B)^{2L}|\psi>|^{1/2L}
\le  \sum_{i=1}^{2M}   
|<\psi^Q|\left(\frac{\partial \hat{B}^{\dagger}}{\partial O_i}\right)^L \left(
\frac{\partial \hat{B}}{\partial O_i}\right)^L |\psi^Q>|^{1/2L} \delta_{i} +...  
\end{equation}
The $n$-order terms (the dots in (14)) are of the general form:
\begin{equation}
\frac{1}{n!}\sum_{i,k...,s=1}^{2M}|<\psi^Q|\left(\frac{\partial^n \hat{B}^{\dagger}}
{\partial O_i \partial O_k...\partial O_s} \right)^L \left( 
\frac{\partial^n \hat{B}}{\partial O_i
\partial O_k...\partial O_s}\right)^L|\psi^Q>|^{1/2L} \delta_{i} \delta_{k}
....\delta_{s}  .
\end{equation} 
This result is valid up to any order since $|\phi^c>$ is $L$-order classical and so the relation $<E_{S^L_{ia}}|E_{S^L_{ia}}>
\le (\delta_{S^L_{ia}})^2$ is valid for all the sequences $S^L_{ia}$ determined in (1)
which are exactly the same ones involved in the expansion of $|<\psi|(\hat A -\hat B)^{2L}|\psi>|$.
This is our last result concerning the value of $|<\psi|(\hat A -\hat B)^{2L}|\psi>|^{1/2L}$. 

To proceed we will construct a general set of states of the form $|\phi^c>|\psi^Q>$ providing a basis to expand the initial data wave function $|\phi>=|\phi^c>|\phi^Q>$ (5).
Let us start by introducing the states $|\psi^r_u>=|\phi^c>|b^r_u>$ where $|b^r_u>$ form a complete set of eigenstates of $\hat B$ with degeneracy index $r$ and associated eigenvalue $b_u$. This would be the most natural set of states to be used to expand $|\phi>$. However, they do not provide suitable results. We will see why in the sequel. To go further we still have to construct another set of 
states and study their properties in the representation of $\hat{A}$. 
Let $|\xi_u>$ be given by:
\begin{equation}
|\xi_u>=|\xi^Q_u>|\phi^c>=
\frac{1}{C_u} \sum_{r,b_u^{\prime} \in I_u}<b^{\prime r}_{u}
|\phi^Q>|b^{\prime r}_{u}>|\phi^c> , 
\end{equation}
where $C_u$ is a normalisation constant, $|b^{\prime r}_u>$ are eigenstates of $\hat B$ and $I_u=[b_u-I_B, b_u+I_B]$ where 
$b_u$ is named the {\it central eigenvalue} associated to $|\xi_u>$ and
$I_B$ is a constant to be supplied later and that represents
the spread of $|\xi_u>$ in the representation of $\hat{B}$. 
We are specially interested in a
set of states $|\xi_u>$ associated to a sequence $S$  
of eigenvalues $b_u$ of $\hat{B}$. These eigenvalues are chosen in such a way 
that their value grows in steps of $2I_B$. This way we guarantee 
that firstly $<\xi_u|\xi_{u^{\prime}}>=\delta_{u,u^{\prime}}$ and 
secondly, that $|\phi>=\sum_{b_u \in S}<\xi_u|\phi>|\xi_u>$.
This said let us study the properties of $|\xi_u>$ in the representation of $\hat{A}$,
(the reader should refer to the appendix for the relevant definitions).\\
{\bf b)} Explicit form of the spread $\Delta_L(\hat{A},\xi_u,b_u,p)$:\\
The $L$-order spread 
$\Delta_L(\hat{A},\xi_u,b_u,p)=<E^L(\hat{A},\xi_u,b_u)|E^L(\hat{A},\xi_u,b_u)>^{1/2L}
/({1-p})^{1/2L}$ can be cast in the form:
\begin{equation}
\Delta_L(\hat{A},\xi_u,b_u,p)=\frac{|<\xi_u|(\hat A - b_u)^{2L}|\xi_u>|^{1/2L}}
{(1-p)^{1/2L}} =
\frac{\left|<\xi_u|\left\{(\hat A - \hat B) +(\hat B - b_u) \right\}^{2L}|\xi_u> \right|^{1/2L}}{(1-p)^{1/2L}}.
\end{equation}
Expanding the polynomial insight the bracket and using the Schwartz inequality $L$ times to separate the terns in $(\hat A - \hat B)$ of those in $(\hat B - b_u)$ we get:
\begin{equation}
\Delta_L(\hat{A},\xi_u,b_u,p) \le 
\frac{1}{(1-p)^{1/2L}}
\left\{ |<\xi_u|(\hat A - \hat B)^{2L}|\xi_u>|^{1/2L} +|<\xi_u|(\hat B - b_u)^{2L}|\xi_u>|^{1/2L} \right\},
\end{equation}
where we disregard the contribution of terms proportional to $\hbar^2$ or smaller.
Using (14) and the inequality
$<E^L(\hat{B},\xi_u,b_u)|E^L(\hat{B},\xi_u,b_u)> \le I_B^{2L}$,
which can easily be obtained from (16), we finally get: 
\begin{equation}
\Delta_L(\hat{A},\xi_u,b_u,p) \le \frac{ 
\sum_{i=1}^{2M} \left| <\xi^Q_u|\left(\frac{\partial \hat{B}^{\dagger}}
{\partial O_i} \right)^L \left(
\frac{\partial \hat{B}}{\partial O_i} \right)^L|\xi_u^Q> \right| ^{1/2L}
\delta_{i} + I_B +....}{(1-p)^{1/2L}}=\frac{\delta_L(\hat B)+I_B}{(1-p)^{1/2L}},
\end{equation}
where $\delta_L(\hat B)$ is named the $L$-order error margin of $\hat B$. The reason for this designation is the resemblance of the expansion (19) and the classical 
error margin for the full classical observable $B=\sum_j B^c_j B_j^Q$:
$\quad \delta_B=\sum_{i=1}^{2M}|\partial B / \partial O_i| \delta_{i}+...$
when the error margins associated to the "quantum" observables are 
identically zero. This resemblance is also clear between the higher order
terms of (19), which are of the form (15), and the higher order terms that are {\it also} present in the expression of the error margin of $B$.  
Finally, notice that if we make $I_B=0$ the set of states $|\xi_u>$ is a set of true eigenvectores of $\hat B$ with associated eigenvalues $b_u$ and with spread $\Delta_L(\hat{A},\xi_u,b_u,p) \le \delta_L(\hat B)/(1-p)^{1/2L}$.

Using the result a) from the appendix we can now state 
that in the representation of $\hat{A}$ 
the state $|\xi_u>$ has at least a probability $p$ in the 
interval $I=[b_u-\Delta_L (p),b_u+\Delta_L (p)]$, with $\Delta_L (p)$ given by (19),
that is $\sum_{n,a_i \in I}|<a_i^n|\xi_u>|^2 \ge p$.

\subsubsection{The initial data wave function in the representation 
of $\hat{A}$}

We have completed all the preliminary work and we are now in position 
to obtain some of the properties of $|\phi>$ in the representation of $\hat{A}$. 
A completely precise prediction for the probability of obtaining the 
eigenvalue $a_i$ from a measurement of $\hat{A}$ given by $P(a_i)=
\sum_{n}|<\phi|a^n_i>|^2$ is not possible due to 
our incomplete knowledge of $|\phi>$. Still, we can attempt to obtain 
fairly accurate predictions.
We will obtain predictions for the
probability of a measurement of $\hat{A}$ yielding   
an eigenvalue $a_i \in I_0$ where $I_0$ is an interval of 
size at least $\Delta_L$, with $\Delta_L$ given by (19). 
More precisely
we will be able to predict that the probability $P(a_i \in I_0)$ is at the
 most $P_{max}(a_i \in I_0)$ and at the least $P_{min}(a_i \in I_0)$, 
the error margin being a function of $L$, typically of reasonable size.

Let us then consider the following three intervals: 
$I_0=[a^0-D,a^0+D]$ is 
the interval of eigenvalues of $\hat{A}$ for which we want 
to determine $P(a_i \in I_0)$. $D$ is required to satisfy $D>\Delta_L$. The two other intervals 
$I_{max}$ and $I_{min}$ will be 
needed to majorate and minorate the former probability. 
$I_{max}$ is such that any $|\xi_u>$ with associated 
central eigenvalue $b_u \in I_0$ has in the representation of $\hat{A}$  
a probability of at least $p$ in $I_{max}$. For $I_{min} $ the 
statement is that if $b_u \in I_{min}$ then $|\xi_u>$ has 
a probability of at least $p$ in $I_0$. Easily we see that 
the intervals $I_{min}=[a^0-(D-\Delta_L),a^0+(D-\Delta_L)]$ and $I_{max}=
[a^0-(D+\Delta_L),a^0+(D+\Delta_L)]$ - where $\Delta_L=\Delta_L(\hat{A},
\xi_u,b_u,p)$ given by (19) and $b_u \in I_0$ - will satisfy the 
former requirements. Notice that we assumed that
$\Delta_L $ has a similar value for different $b_u$
within $I_0$. If this is not the case all the future results 
are still valid, we just need to be more careful in constructing 
the intervals $I_{max}$ and $I_{min}$.  
Let us then proceed. We have:
\begin{equation}
P(a_i \in I_0)=\sum_{n,a_i \in I_0} |<\phi|a_i^n>|^2=
\sum_{n, a_i \in I_0} \left|\sum_{b_u \in S}<\phi|\xi_u><\xi_u|a^n_i> 
\right|^2 , 
\end{equation}
where we have used the fact that $|\phi>=\sum_{b_u \in S} <\xi_u|\phi>|\xi_u>$,
where $S$ is the set of central eigenvalues of $\hat{B}$, associated to the states $|\xi_u>$, that was presented in
the sequel of (16).
We now expand the previous expression first using the interval
$I_{max}$:
\begin{eqnarray}
P(a_i \in I_0) & = &
\sum_{n, a_i \in I_0} \left|\sum_{b_u \in I_{max} \cap S}<\phi|\xi_u><\xi_u|a^n_i> 
+\sum_{b_u \in S/I_{max}}<\phi|\xi_u><\xi_u|a^n_i> 
\right|^2 \nonumber\\
& \le &  \sum_{n, a_i} \left|\sum_{b_u \in I_{max} \cap S}<\phi|\xi_u><\xi_u|a^n_i> 
\right|^2 + \sum_{n, a_i \in I_0} \left|\sum_{b_u \in S/I_{max}}
<\phi|\xi_u><\xi_u|a^n_i> \right|^2 \nonumber\\
& + & 2 \left| \sum_{n, a_i \in I_0} \sum_{b_u \in I_{max} \cap S}
<\phi|\xi_u><\xi_u|a^n_i> \sum_{b_u \in S/I_{max}}
<a^n_i|\xi_u><\xi_u|\phi> \right| ,
\end{eqnarray}
where the set $S/I_{max}$ is constituted by the elements of $S$ that do 
not belong to $I_{max}$. To derive the former expression we first wrote
the norm in the first summation as a product of the term inside the norm
by its complex conjugate and then grouped the resulting terms in
a convenient way.

On the other hand, and using the interval $I_{min}$ we have: $P(a_i \in I_0)=
1-P(a_i \notin I_0)$ and so:
\begin{eqnarray}
P(a_i \in I_0) & = &  
1-\sum_{n, a_i \notin I_0} \left|\sum_{b_u \in S/I_{min}}<\phi|\xi_u><\xi_u|a^n_i> 
+\sum_{b_u \in I_{min} \cap S}<\phi|\xi_u><\xi_u|a^n_i> 
\right|^2 \nonumber\\
& \ge & 1-\sum_{n, a_i} \left|\sum_{b_u \in S/I_{min}}<\phi|\xi_u><\xi_u|a^n_i> 
\right|^2 - \sum_{n, a_i \notin I_0} \left|\sum_{b_u \in I_{min} \cap S}
<\phi|\xi_u><\xi_u|a^n_i> \right|^2 \nonumber\\
& - & 2 \left|\sum_{n, a_i \notin I_0} \sum_{b_u \in S/I_{min}}
<\phi|\xi_u><\xi_u|a^n_i> \sum_{b_u \in I_{min} \cap S}
<a^n_i|\xi_u><\xi_u|\phi> \right| .
\end{eqnarray}
The right hand side of the former inequalities has three and four 
terms, respectively.
Let us designate by $X_1,Z_1$ the second and third terms of 
(21) and by $X_2,Z_2$ the third and fourth terms of (22). 
We will deal with each of the terms in (21) and (22) independently: \\
{\bf a)} First term of (21):
\begin{eqnarray}
\sum_{n, a_i} \left|\sum_{b_u \in I_{max} \cap S}<\phi|\xi_u><\xi_u|a^n_i> 
\right|^2 & = &
\sum_{n, a_i} \sum_{b_u,b_v \in I_{max} \cap S}<\phi|\xi_u><\xi_v|\phi>
<\xi_u|a^n_i><a^n_i|\xi_v> \nonumber\\
& = & \sum_{b_u \in I_{max} \cap S} |<\phi|\xi_u>|^2 = P(b_u \in I_{max}),
\end{eqnarray}
since $\sum_{n, a_i} <\xi_u|a^n_i><a^n_i|\xi_v> =\delta_{uv}$ and
 $\sum_{b_u \in I_{max} \cap S} |<\phi|\xi_u>|^2=
\sum_{r,b_u \in I_{max}} |<\phi|\psi^r_u>|^2$ where $|\psi^r_u>$ are 
the eigenstates of $\hat{B}$.\\ 
{\bf b)} First terms of (22): Just as for the previous term we have:
\begin{equation}
1-\sum_{n, a_i} \left|\sum_{b_u \in S/I_{min}}<\phi|\xi_u><\xi_u|a^n_i> 
\right|^2 =1-P(b_u \notin I_{min})=P(b_u \in I_{min}).
\end{equation}
{\bf c)} Third term of (21) and fourth term of (22):
\begin{eqnarray}
& & Z_1  =  2 \left| \sum_{n, a_i \in I_0} \sum_{b_u \in I_{max} \cap S}
<\phi|\xi_u><\xi_u|a^n_i> \sum_{b_u \in S/I_{max}}
<a_i^n|\xi_u><\xi_u|\phi> \right| \\ 
& = & 2\left|\sum_{n, a_i \in I_0} \sum_{b_u \in I_{max} \cap S}
<\phi|\xi_u><\xi_u|a^n_i> <a^n_i| \cdot
\sum_{m, a_j \in I_0} \sum_{b_v \in S/I_{max}}
<a^m_j|\xi_v><\xi_v|\phi> |a^m_j>\right| ,\nonumber
\end{eqnarray}
using the Shwartz inequality to calculate the former inner product
and taking result a) into account we obtain:
\begin{equation}
Z_1 \le 2P(b_u\in I_{max})^{1/2}
\left(\sum_{n, a_i \in I_0} \left|\sum_{b_u \in S/I_{max}}
<\phi|\xi_u><\xi_u|a^n_i> \right|^2 \right)^{1/2},
\end{equation}
where the second term of the former product is exactly $X^{1/2}_1$ where
$X_1$ is the second term in (21) to be calculated in d).

Following exactly the same procedure we get for the last term in (22):
\begin{equation}
Z_2 \ge -2 P(b_u \notin I_{min})^{1/2}
\left(\sum_{n, a_i \notin I_0} \left|\sum_{b_u \in I_{min} \cap S}
<\phi|\xi_u><\xi_u|a^n_i> \right|^2 \right)^{1/2},
\end{equation}
where the second term of the right hand side of the former inequality
is exactly $-X^{1/2}_2$ where $X_2$ is the third term of (22) 
to be calculated in d). \\
{\bf d)} Second term of (21) and third term of (22): 
These are the last terms we need to calculate in order to obtain 
the explicit form of the expressions (21) and (22). Let us start 
with $X_1$:
\begin{eqnarray}
X_1 &=& \sum_{n, a_i \in I_0} \left|\sum_{b_u \in S/I_{max}}
<\phi|\xi_u><\xi_u|a^n_i> \right|^2 \nonumber\\ 
&\le & \sum_{b_u,b_v \in S/I_{max}}\left|<\phi|\xi_u><\xi_v|\phi>\right|
\left|\sum_{n, a_i \in I_0} <\xi_u|a^n_i><a^n_i|\xi_v>\right|.
\end{eqnarray}
For the second term of the right hand side of
the previous inequality, we get:
\begin{equation}
\left|\sum_{n, a_i \in I_0} <\xi_u|a^n_i><a^n_i|\xi_v>\right|
\le \left(\sum_{n, a_i \in I_0} |<\xi_u|a^n_i>|^2\right)^{1/2}
\left(\sum_{n, a_i \in I_0} |<\xi_v|a^n_i>|^2\right)^{1/2},
\end{equation}
where we have used the Shwartz inequality to calculate
the inner product of the states $ \sum_{n, a_i \in I_0} <\xi_u|a^n_i><a^n_i|$
and $\sum_{n, a_i \in I_0}<a^n_i|\xi_v> |a^n_i>$.  
To proceed we notice that both $b_u, b_v \notin I_{max}$ and  
use the result (67) in b) from the appendix to get:
\begin{equation}
\sum_{n, a_i \in I_0} |<\xi_u|a^n_i>|^2 \le (1-p) \frac{\Delta_L (\hat{A},\xi_u ,b_u,p)^{2L}}{|b_u-a|^{2L}},
\end{equation}
where $\Delta_L (\hat{A},\xi_u ,b_u,p)$ is given by (19) and $a$ is one of the extremes of the interval $I_0$, the one that minimise the distance $|b_u-a|$, that is $a=a^0+D$ or 
$a=a^0-D$. 
Putting these results together we get:
\begin{equation}
X_1 \le (1-p)
\left\{\sum_{b_u \in S/I_{max}}\frac{\Delta_L (\hat{A},\xi_u ,b_u,p)^L
|<\phi|\xi_u>|}{|b_u-a|^L}\right\}^2.
\end{equation}
Since $|<\phi|\xi_u>|=|<\phi^Q|\xi_u^Q>|$ 
the calculation of the right hand side of inequality (31) might be done explicitly 
once the initial data of the half quantum system and the operator $\hat B$ are given. Therefore, we can now group the former results together to obtain a prediction for $P(a_i \in I_0)$. However, and 
since we would like to obtain a more explicit value for $P(a_i \in I_0)$, let us consider the simple but quite general case in which
$\Delta_L (\hat{A},\xi_u ,b_u,p)$ is approximately a constant in the range of eigenvalues where $|<\phi|\xi_u>|$
have meaningful values. In this case we have:  
\begin{equation}
X_1 \le (1-p)\Delta_L(p)^{2L}
\left\{\sum_{b_u \in S/I_{max}}\frac{|<\phi|\xi_u>|}{|b_u-a|^L}
\right\}^2.
\end{equation}
The goal is to maximize the term inside the brackets to obtain the highest possible value of $X_1$. Let then $|\phi>=\sum_{b_u \in S}<\xi_u|\phi>|\xi_u>$ and let $C_u=|<\xi_u|\phi>|$.
As an intermediate step let us assume that $|\phi>$ spreads for an interval from $E_0=a^0+\Delta_L +D$ (the extreme of $I_{max}$) to an arbitrary $E \in {\cal R}$, i.e. $C_u=0$ for $b_u \notin [E_0,E]$. In the end we will see that the result for $X_1$ is independent of $E$ that might be taken to infinity.  
Let us proceed: since $b_u$ grows in steps of size $2I_B$,
we divide the former interval in sub-intervals of size $2I_B$.
Let us say that we have $N$ such sub-intervals: $2I_B=(E-E_0)/N$.
We then get:
\begin{equation}
\sum_{b_u \in S/I_{max}} \frac{|<\phi|\xi_u>|}{|b_u-a|^L}=
\sum_{n=0}^N \frac{C_n}{(\Delta_L+2nI_B)^L}=
\frac{1}{2I_B} \sum_{n=0}^N 2I_B 
\frac{C_n}{(\Delta_L+2nI_B)^L}
\simeq \frac{1}{2I_B} \int_{E_0}^E \frac{C(x)}{|x-a|^L}dx ,
\end{equation}
where $C_n=|<\phi|\xi_n>|$ with $|\xi_n>$ being the state associated 
to the central eigenvalue $b_n=a+\Delta_L+2nI_B$. Our task now is to maximise
the previous integral subject to the constraint:
\begin{equation}
\sum_{n=0}^N C^2_n =1 \Longrightarrow \frac{1}{2I_B}
\int_{L_0}^L C(x)^2 d x =1.
\end{equation}
To do this we use the Lagrangian multiplier method. We put
$L(C,\dot{C},x)=-C/|x-a|^L + \lambda C^2$
and we write the Lagrangian equations to obtain:
\begin{equation}
\frac{\partial L}{\partial C}-\frac{\partial}{\partial x}
\frac{\partial L}{\partial \dot{C}}=0 \quad \Longleftrightarrow \quad
C(x)=\frac{1}{2\lambda|x-a|^L}.
\end{equation}
And imposing the constraint (34) in $C(x)$ we get:
\begin{equation}
\lambda=\left\{ \frac{1}{8(2L-1)I_B} \left( \frac{1}{(E_0-a)^{2L-1}}-\frac{1}{(E-a)^{2L-1}}
\right) \right\}^{1/2},
\end{equation}
substituting this result in (35), integrating (33) and, finally, substituting the value of this integral in (32) we get:
\begin{equation}
X_1 \le (1-p)\frac{\Delta_L(p)}{2(2L-1)I_B},
\end{equation}
which is our final result for $X_1$. We see that this result could not be obtained if we had used the eigenstates $|\psi^r_u>$ of $\hat B$ in which case $I_B=0$ and thus $X_1$ will not be bound.

If we proceed just along the same lines for the third term in (22) we will
get exactly the same result, that is:
\begin{equation}
X_2 = -\sum_{n, a_i \notin I_0} \left|\sum_{b_u \in I_{min} \cap S}
<\phi|\xi_u><\xi_u|a^n_i> \right|^2 
\ge - (1-p)\frac{\Delta_L(p)}{2(2L-1)I_B}.
\end{equation}
We now introduce the results of a),b),c),d) into (21) and (22) and finally get:
\begin{equation}
P(b_u \in I_{min})-E_{min}
\le P(a_i \in I_0) 
\le P(b_u \in I_{max})+E_{max},
\end{equation}
where $P(b_u \in I_{min})= \sum_{r,b_u \in I_{min}}|<b_u^r|\phi^Q>|^2$,
$P(b_u \in I_{max})= \sum_{r,b_u \in I_{max}}|<b_u^r|\phi^Q>|^2$ and
$E_{min}$ and $E_{max}$ are given by the following expressions:
\begin{eqnarray}
E_{min}=2P(b_u \notin I_{min})^{1/2} \left(\frac{(1-p) \Delta_L(p)}
{2(2L-1)I_B}\right)^{1/2} + \frac{(1-p) \Delta_L(p)}{2(2L-1)I_B}
\nonumber\\
E_{max}=2P(b_u \in I_{max})^{1/2} \left(\frac{(1-p) \Delta_L(p)}
{2(2L-1)I_B}\right)^{1/2} + \frac{(1-p) \Delta_L(p)}{2(2L-1)I_B} .
\end{eqnarray}
Notice that, given the degree of validity $L$ of the classical sector initial data, we can play with the interval $I_0$ and with the values of 
$I_B$ and $p$, which
in turn impose a value to $\Delta_L(p)$, to minimise the error of the 
predictions for the probabilities. 
To get some feeling about the accuracy of the predictions let us choose some explicit
values for $L, I_B$ and $p$. Let $I_B = \delta_L(\hat B)$ so that $\Delta_L(p)=2I_B/ (1-p)^{1/2L}$. The errors $E_{min}$ and $E_{max}$ become: 
\begin{eqnarray}
E_{min}=2P(b_u \notin I_{min})^{1/2} \left(\frac{(1-p)^{\frac{2L-1}{2L}}}
{2L-1}\right)^{1/2} + \frac{(1-p)^{\frac{2L-1}{2L}}}{2L-1}
\nonumber\\
E_{max}=2P(b_u \in I_{max})^{1/2} \left(\frac{(1-p)^{\frac{2L-1}{2L}}}
{2L-1}\right)^{1/2} + \frac{(1-p)^{\frac{2L-1}{2L}}}{2L-1}.
\end{eqnarray}
Let us consider $L=1$. This means that the classical sector initial data $(O^0_i,\delta_i)$ is first-order valid, i.e. the classical sector initial data wave function $|\phi^c>$ is first-order classical with respect to the classical data $(O^0_i,\delta_i)$. Let us also choose $p=0.99$. We can then state that, in the representation of $\hat A$, the states $|\xi_u>$ (16) have, at least, $99\%$ of their probability confined to the intervals: 
$I_u=[b_u-20\delta_1(\hat B),b_u+20 \delta_1(\hat B)]$ where $\delta_1(\hat B)$ is given by
(19) and is of the size of a classical error margin. Moreover:
\begin{equation}
E_{min}=2\times 0.31 \times P(b_u \notin I_{min})^{1/2}  + 0.1 \quad {\rm and} \quad
E_{max}=2 \times 0.31 \times P(b_u \in I_{max})^{1/2} +0.1 ,
\end{equation}
and thus, in the worst case:
\begin{equation}
P(b_u \in I_{min})-0.72
\le P(a_i \in I_0) 
\le P(b_u \in I_{max})+0.72 ,
\end{equation}
with the difference between the ranges of $I_{max},I_{min}$ and $I_0$ being given by $20 \delta_1(\hat B)$. An error of $72\%$ is huge. The reason for such a large error lies upon the fact that the conditions imposed over $|\phi^c>$ are the least restrictive possible, $L=1$. In other words, the classical initial data is the least valid possible.

To see what happens when we increase the degree of validity of the classical sector initial data let us make $L=10$. We consider once again $I_B=\delta_{10}(\hat B)$ but, this time let us choose $p=0.99999 =1-10^{-5}$. We have $\Delta_{10}(p=0.99999)=3.6 \delta_{10}(\hat B)$ and:
\begin{equation}
E_{min}=0.0019 P(b_u \notin I_{min})^{1/2}  + 9.4 \times 10^{-7} , \quad
E_{max}=0.0019 P(b_u \in I_{max})^{1/2} + 9.4 \times 10^{-7} ,
\end{equation}
and thus, in the worst case:
\begin{equation}
P(b_u \in I_{min})-0.0019
\le P(a_i \in I_0) 
\le P(b_u \in I_{max})+0.0019.
\end{equation}
That is an error of $0.19\%$ with the difference between the ranges of $I_{max},I_{min}$ and $I_0$ decreasing to $3.6 \delta_{10}(\hat B)$.

\section{Half quantization}

We start by noticing that the predictions $P(a_i \in I_0) $ and its 
error margins could be obtained if we had knowledge of 
the operator $\hat{B}$ and in no way require (except to obtain 
$\hat{B}$) the knowledge of the full quantum operator $\hat{A}$.
This means that if we were able to calculate $\hat{B}$ directly
then we would be able to make predictions for the evolution of the 
half quantum system without firstly having to obtain its full quantum
version. 
Therefore, the aim of this section is to obtain a framework able to provide the 
operator $\hat{B}$ directly from the initial data of the half quantum  
system without requiring previous knowledge of the full quantum theory.

\subsection{The unquantization map}

In section 2.2 we present the first definition of the unquantization map. 
The motivation to define $V_0$ this way was the fact that it validates the expansions (10,11)
which were crucial to develop the entire approximation procedure presented in the last section.
It was already pointed out that the map $V_0$ is a trivial generalisation of the 
unquantization map presented in \cite{nuno1}. Let us name this last map $V_0^c$. In fact the action of $V_0$ over 
a classical sector operator is identical to the action of $V_0^c$. In that paper we saw that $V_0^c$ is just the inverse map of the Dirac quantization \cite{dirac1, dirac2}. Taking this result into account  we present a new, however equivalent, definition of the half unquantization:

\bigskip
  
\underline{{\bf Definition II}} {\bf - Second unquantization map}\\
Let $\wedge$ be the Dirac quantization map \cite{dirac2}, $\wedge:{\cal A}(T^{\ast}M_1) \longrightarrow {\cal A}({\cal H}_1)$ and let $\hat{A}=
\sum_j \hat{A}_{j}^Q \hat{A}_j^c$ where $\hat{A}_j^Q$ and $\hat{A}^c_j$ are arbitrary multiple products of quantum and classical sector operators, respectively (7). 
The unquantization $\vee$ is a map from the algebra ${\cal A}({\cal H})$ 
to the algebra ${\cal S}$ (check for the definition of ${\cal S}$ in {\it definition I}) defined by the following rules: \\
{\bf 1)}
$\vee (\hat{A})= \sum_j \vee (\hat{A}_{j}^Q) \vee (\hat{A}_j^c)$\\
{\bf 2)}
$\vee(\hat{A}_j^c)=A_i^c \qquad \mbox{iff} \qquad \wedge(A_j^c)=\hat{A}_j^c$\\
{\bf 3)} 
$\vee(\hat{A}_j^Q)=\hat{A}_j^Q$

Let us study some properties of $\vee$:\\
1) The map $\vee$ is equivalent to the map $V_0$ of definition I. This is so because the rule 2) of the definition of $\vee$ is equivalent to the rule 3) of the definition of $V_0$. This fact was extensively discussed in \cite{nuno1}.\\
2) Just like $V_0$, the map $\vee$ is beset by order problems. In general there are several different classical sector observables that when quantized yield the same quantum operator. Let $A_1^c \not= A_2^c$ be two such observables, i.e. $\wedge (A^c_1) = \hat{A}^c$ and also $\wedge (A^c_2) = \hat{A}^c$. This means that $\vee(\hat{A}^c)=A^c_1$ but also $\vee (\hat{A}^c)=A^c_2$. Hence the map $\vee$ is not univocous. 
On the other hand, the predictions (39,40) of the last section, are made for a general quantum operator $\hat{A}$ (for instance $\hat{A}=\hat{A}^c\hat{A}^Q$) and might be obtained using any of the operators $\hat{B}=V_0(\hat{A})$ (or equivalently, $\hat{B}=\vee(\hat{A})$).
Therefore, the ambiguity of $\vee$ could be problematic if the predictions obtained by using two different $\hat{B}$ (for instance $\hat{B}_1={A}_1^c\hat{A}^Q$ and $\hat{B}_2={A}_2^c\hat{A}^Q$) were inconsistent.

However, one can easily realise that this is not the case. The difference $\hat{B}_2-\hat{B}_1$ is proportional to a leading factor of $c_j\hbar^2$ (where $c_j$ is the highest numerical coefficient of $\hat{A}$ displayed in the orders from which $\hat{B}_1$ and $\hat{B}_2$ were calculated). We already saw in the sequel of (11) that the validity of the predictions of the last section rests upon the premise that the numerical factors of $\hat{B}>>c_j\hbar^2$ (otherwise $\hat{B}$ can not be considered for reproducing the predictions of $\hat{A}$). We also saw that this premise is satisfied if the original operator $\hat{A}$, from which $\hat{B}$ was calculated, satisfies some order conditions. Therefore, and if $\hat{B}_2$ and $\hat{B}_1$ are both valid operators, obtained from $\hat{A}$ displayed in required orders, the difference $\hat{B}_2-\hat{B}_1$ is not meaningful when compared to the imprecision (which is proportional to the numerical factors of $\hat B$ (19)) associated to the predictions obtained by using either $\hat{B}_1$ or $\hat{B}_2$. 
In conclusion,
$\hat{B}_1$ and $\hat{B}_2$ provide {\it physical predictions} which are consistent with each other, solving the ambiguity. \\
3) Unquantizing of the product of two classical sector operators: let us consider two general classical observables $B$ and $C$. To quantize $BC$ one uses the 
symmetrization rule: $\wedge (BC)=1/2 (\hat{B}\hat{C}+\hat{C}\hat{B})$.
We just use the same rule for the unquantization:
\begin{equation}
\vee(\hat{B}\hat{C})= \vee \left( \frac{\hat{B}\hat{C}+\hat{C}\hat{B}}{2} +
\frac{1}{2} [\hat{B},\hat{C}] \right) =BC+\frac{1}{2}i\hbar \{B,C\} .
\end{equation}
Notice that the prescription is beset by order problems (comment 2).\\
4) Unquantization of a self-adjoint operator:
if $\hat{A}=\hat{A}^c$ then we get from rule 2):
$\vee(\hat{A}^{\dagger})=\vee(\hat{A})^{\ast}$. For the case of a 
general operator $\hat{A}=\sum_j \hat{A}^c_j \hat{A}^Q_j$ we have:
\begin{equation}
\vee(\hat{A}^{\dagger})=\sum_j \vee(\hat A^c_j)^{\ast}({\hat{A}^Q_j})^{\dagger}=
\vee(\hat{A})^{\dagger},
\end{equation}
and if $\hat{A}=\hat{A}^{\dagger}$ then $\vee(\hat{A})^{\dagger}=
\vee(\hat{A}^{\dagger})=\vee(\hat{A})$ and so $\vee(\hat{A})$ is 
also self-adjoint.\\
5) Unquantizing the brackets:
For the simplest case of $ \hat{A}=\hat{A}^c$ and 
 $\hat{B}=\hat{B}^c$, from rule 2) one immediately has:
\begin{equation}
\vee [\hat{A},\hat{B}]=\vee \left(\wedge(i\hbar\{A,B\})\right)=i\hbar\{A,B\}.
\end{equation}
For the most general case let us first put $\hat{A}=\hat{A}^c\hat{A}^Q$ and 
$\hat{B}=\hat{B}^c\hat{B}^Q$ which only excludes sums of operators which, using
rule 1), are straightforward to handle.
We get:
\begin{eqnarray}
[\hat{A},\hat{B}] & = & \hat{A}^c \hat{B}^c[\hat{A}^Q,\hat{B}^Q]+
[\hat{A}^c,\hat{B}^c] \hat{B}^Q \hat{A}^Q \nonumber \\  
\Longrightarrow 
\vee([\hat{A},\hat{B}]) & = & \vee(\hat{A}^c \hat{B}^c)[\hat{A}^Q,\hat{B}^Q]+
\vee([\hat{A}^c,\hat{B}^c]) \hat{B}^Q \hat{A}^Q ,
\end{eqnarray}
and using (46) and (48) we get:
\begin{equation}
\vee([\hat{A},\hat{B}])  =  A^c B^c [\hat{A}^Q,\hat{B}^Q]+
\frac{i\hbar}{2} \{A^c,B^c\} ( \hat{A}^Q \hat{B}^Q + \hat{B}^Q \hat{A}^Q ).
\end{equation}

\subsection{Half quantization and half quantum mechanics}

Eq.(50) can be displayed in a slightly different form:
\begin{equation}
\vee([\hat{A},\hat{B}])  =  [\vee(\hat{A}),\vee(\hat{B})]+
i\hbar \{\{\vee(\hat{A}),\vee(\hat{B})\}\} = (\tilde A,\tilde B),
\end{equation}
where the double brackets are defined by:
\begin{eqnarray}
\{\{\vee(\hat{A}),\vee(\hat{B})\}\} & = &
\frac{1}{2} \{A^c,B^c\} ( \hat{A}^Q \hat{B}^Q + \hat{B}^Q \hat{A}^Q ) 
\nonumber \\
 & = & \frac{1}{2} \sum_i \frac{\partial \tilde{A}}{\partial q_i} 
\frac{\partial \tilde{B}}{\partial p_i}- \frac{\partial \tilde{A}}
{\partial p_i} 
\frac{\partial \tilde{B}}{\partial q_i} + \frac{\partial \tilde{B}}
{\partial p_i} 
\frac{\partial \tilde{A}}{\partial q_i} - \frac{\partial \tilde{B}}
{\partial q_i} 
\frac{\partial \tilde{A}}{\partial p_i},
\end{eqnarray}
and we introduced the notation 
$\tilde{A}=\vee(\hat{A})$ and defined the new bracket $(\quad ,\quad )=[\quad ,\quad ] +
i\hbar \{\{\quad ,\quad \}\}$.
This bracket was first proposed in \cite{aleksandrov,boucher}. The motivation to define it this way 
was given in terms of the properties of the emerging theory, namely that it properly generalises 
both quantum and classical mechanics. The bracket is known to be antisymmetric, multilinear
but it does not satisfy the Jacobi identity. This did cause much debate in the literature \cite{anderson1,diosi1,jones,salcedo1,salcedo2}. We will come back to this problem in the conclusions. Firstly let us finish the presentation of the dynamical structure of half quantum mechanics.
 
Given the unquantization map of definition II we are able to define a quantization prescription mapping a general classical dynamical system to the correspondent half quantum one.
One just needs to specify the classical and the quantum sectors to be of the original classical theory, that
is to provide $T^{\ast}M_1$ and $T^{\ast}M_2$.
Let then ${\cal F}$ be the algebra of observables over $T^{\ast}M=T^{\ast}M_1 \otimes T^{\ast}M_2$. The 
remaining notation is in accordance with the previous definition. 

\bigskip

\underline{{\bf Definition III}} {\bf - Half quantization map}\\
The half quantization is defined to be the map:
\begin{equation}
\cap :{\cal F} \longrightarrow {\cal S}; \quad \cap=\vee \circ 
\wedge \quad A \longrightarrow \tilde{A}=\cap(A),
\end{equation}
where $\wedge : {\cal F} \longrightarrow {\cal A}({\cal H}_1 \otimes {\cal H}_2) $ is the Dirac quantization map and $\vee$ is the unquantization map of definition II.

The properties of $\bigcap$ follow directly from its definition:
Let $A,B \in {\cal F}$:\\
{\bf 1)} $
\cap \left( f(q_i,p_i) \right)=f(q_i,p_i)\hat{I}$ and $
\cap (q_{\alpha}) = \hat{q}_{\alpha}$ , $
\cap (p_{\alpha}) = \hat{p}_{\alpha}$, $i=1..M$, $\alpha=(M+1)..(M+N)$\\
{\bf 2)} $\cap$ is a linear map: $\cap (A+bB)=\cap(A) + b\cap(B)$, $b \in {\cal C}$\\
{\bf 3)} $\cap(f(q_i,p_i)g(q_{\alpha},p_{\alpha}))=f(q_i,p_i)
\wedge(g(q_{\alpha},p_{\alpha})) $\\
{\bf 4)}
$\cap (\{A,B\})=(i\hbar)^{-1} {\bf (}\cap(A),\cap(B) {\bf )}$\\

We are now in position to study the theory resulting from applying the 
half quantization procedure to a given classical theory. 
First we have to choose a CSCO for
the quantum sector of the theory, let it be, for instance, the set 
$\{\hat{q}_{\alpha}\},\alpha =M+1..M+N$. 
The initial data for the classical sector is given by the set: 
$\{q_i^0=q_i(t_0),p_i^0=p_i(t_0)\}$ 
and the correspondent error margins $\delta_{q_i}$
and $\delta_{p_i},i=1...M$. The quantum initial data is given  
by the initial data wave 
function $|\phi^Q> \in {\cal H}_2$.
The dynamical evolution of the half quantum system is determined by the 
following set of equations:
\begin{equation}
\dot{\tilde{O}}_{k} = \cap(\{O_k,H\})
 =  \frac{1}{i\hbar} {\bf (}\tilde{O}_{k},\tilde{H}{\bf )} , 
\end{equation}
where $\tilde O_k,k=1...2(M+N)$ is any of the fundamental variables.
The former set of equations has the formal solutions:
\begin{equation} 
\tilde{O}_{k}(t)  =  \sum_{n=0}^{\infty} \frac{1}{n!}
\left(\frac{t}{i\hbar}\right)^n
{\bf (}...{\bf (}\tilde{O}_{k},\tilde{H}{\bf )}...,\tilde{H}{\bf )}
=O_k(\hat{q}_{\alpha},\hat{p}_{\alpha},q_i^0,p_i^0,t).
\end{equation}
Notice that (54) is just the same set of equations as the one 
resulting from applying the unquantization map to the standard quantum evolution 
equations for the observables $\hat{O}_{k}(t)$ and so is the solution 
(55). Hence the observables $\tilde O_k(t)$ are just the operators $\hat B$ we need to supply to obtain the predictions (39,40).

\section{Example}

To illustrate the procedure by which half quantum mechanics makes predictions for the time evolution of a given dynamical system let us consider the following system of two interacting particles described by the Hamiltonian:
\begin{equation}
\tilde H= \frac{\hat P^2}{2M}+\frac{p^2}{2m} + kq\hat P,
\end{equation}
where $(\hat Q,\hat P)$ are the fundamental observables of the quantum particle of mass $M$, 
$(q,p)$ are the canonical variables of the classical particle of mass $m$ and $k$ is a coupling constant. The initial time configuration of the quantum particle is described 
 by the quantum sector wave function $|\phi^Q>$, while the initial time configuration of the classical particle is described by the data $\{q(0),p(0),\delta_q(0),\delta_p(0)\}$.

Solving the half quantum equations of motion (54) we obtain the time evolution of the fundamental observables of the half quantum system, together with the errors $\delta_L(\hat B)$ of the half quantum operators (19):
\begin{equation}
\left\{ \begin{array}{lll}
\tilde q(t) & = & q(0)+ \frac{p(0)}{m} t -\frac{k \hat P(0)}{2m} t^2 \\
\tilde p(t)  & = & p(0) - k \hat P(0) t \\
\tilde Q(t) & = & \hat Q(0) + \left( \frac{\hat P(0)}{M} +kq(0) \right) t \\
&& + \frac{k}{2m} p(0) t^2 -\frac{k^2}{6m}\hat P(0) t^3\\
\tilde P(t) & = & \hat P(0)
\end{array} \right. \qquad , \qquad 
\left\{ \begin{array}{l}
\delta_L(\tilde q(t))  =   \delta_q(0) + |\frac{t}{m}| \delta_p(0) \\
\delta_L(\tilde p(t))  =   \delta_p(0) \\
\delta_L(\tilde Q(t))  =   |kt|\delta_q(0) +|\frac{kt^2}{2m}|\delta_p(0) \\
\delta_L(\tilde P(t))  =    0 
\end{array} \right. ,
\end{equation}
a result that is valid for all $L \in {\cal N}$. The spreads are of the general form
$\Delta_L(\hat B)= \delta_L(\hat B) /(1-p)^{1/2L}$, for $\hat B=\tilde q(t),\tilde p(t), \tilde Q(t)$ or $\tilde P(t)$.

Let $\hat{A}$ be one of the full quantum operators $\hat q(t),\hat p(t), \hat Q(t)$ or $\hat P(t)$, the ones we would have obtained if we had performed the full quantum treatment of the system with Hamiltonian $\hat H=\hat P^2/2M +\hat p^2/2m +k\hat q \hat P$. Let $\hat B$ be the correspondent half quantum operator (57). Moreover, let $|a_i^n>$ be a complete set of eigenstates of $\hat A$ ($a_i$ is the associated eigenvalue and $n$ is the degeneracy index)
and $|b_u^r>$ a complete set of eigenvectores of $\hat B$ ($b_u$ is the associated eigenvalue and $r$ is the degeneracy index). If the classical sector initial data is taken to be first-order valid ($L=1$) the half quantum predictions for the outputs of a measurement of the full quantum operator $\hat A$ (choosing $p=0.99$ and $I_B=\delta_1(\hat B)$) are given by (43):
$$
P(b_u \in I_{min})-0.72 \le P(a_i \in I_0) \le P(b_u \in I_{max})+0.72,
$$ 
where $I_0=[a^0-D,a^0+D]$ is an arbitrary interval centred at $a^0\in {\cal R}$ with range $D\ge 20 \delta_1(\hat B)$, the error $\delta_1(\hat B)$ is given by (57), $I_{max}=[a^0-(D+20\delta_1(\hat B)),a^0+(D+20\delta_1(\hat B))]$ and $I_{min}=[a^0-(D-20\delta_1(\hat B)),a^0+(D-20\delta_1(\hat B))]$. Moreover, $P(b_u \in I_{min,max})=\sum_{r,b_u\in I_{min,max}}|<\phi^Q|b_u^r>|^2$.

If the classical sector initial data $\{q(0),p(0),\delta_q(0),\delta_p(0)\}$ is 10th-order valid then, as we have seen, the precision of the half quantum predictions increases considerably (let $p=0.99999$ and  $I_B=\delta_{10}(\hat B)$):
$$
P(b_u \in I_{min})-0.0019 \le P(a_i \in I_0) \le P(b_u \in I_{max})+0.0019,
$$ 
where, this time, $I_{max}=[a^0-(D+3.6\delta_{10}(\hat B)),a^0+(D+3.6\delta_{10}(\hat B))]$, $I_{min}=[a^0-(D-3.6\delta_{10}(\hat B)),a^0+(D-3.6\delta_{10}(\hat B))]$ and $\delta_{10}(\hat B) =\delta_1(\hat B)$ is given by (57).

Clearly the former predictions are not valid in general. They are valid if the two descriptions of the classical sector initial time configuration, the classical 
$\{q(0),p(0),\delta_q(0),\delta_p(0)\}$ and the quantum $|\phi^c>$, satisfy some consistency conditions.
Given the classical initial data let us see what are the wave functions $|\phi^c>$ that satisfy the $L$-order classicality conditions.

Following the procedure of section 2.1, 
the $L=1$ fundamental sequences (1) are:  
\begin{equation}
S_1=q \quad {\rm and} \quad S_2 = p,
\end{equation}
and the $L$-order sequences:
\begin{equation}
S^{(L)}=(z_1,...,z_L), \quad z_i = q \vee p, \quad i=1..L,
\end{equation}
and thus the $L$-order classicality condition (3) reads:
\begin{eqnarray}
&& <E_{S^{(L)}}|E_{S^{(L)}}> \le \delta^2_{S^{(L)}} \quad , \forall S^{(L)} \quad {\rm in \quad (59)}\\
&& \Longleftrightarrow \quad
<\phi^c|(\hat{z}_L-z_L(0))...(\hat{z}_1-z_1(0))(\hat{z}_1-z_1(0))...(\hat{z}_L-z_L(0))
|\phi^c> \le \delta_{z_1}(0)^2... \delta_{z_L}(0)^2 , \nonumber
\end{eqnarray}
Using the Shwartz inequality and disregarding the contribution of terms proportional to $\hbar^2$, the former inequalities are reduced to:
\begin{equation}
\left\{ \begin{array}{l}
 <\phi^c|(\hat q-q(0))^{2L}|\phi^c> \le \delta_q(0)^{2L}\\
\\
 <\phi^c|(\hat p-p(0))^{2L}|\phi^c> \le \delta_p(0)^{2L}
\end{array} \right.
\quad \Longleftrightarrow  \quad
\left\{ \begin{array}{l}
\int (q-q(0))^{2L} |\phi^c (q)|^2 dq  \le  \delta_q(0)^{2L} \\
\\
\int (p-p(0))^{2L} |\phi^c (p)|^2 dp  \le  \delta_p(0)^{2L}
\end{array} \right. 
\end{equation}
Given the classical initial data $\{q(0),p(0),\delta_q(0),\delta_p(0)\}$, (61) constitute a system of inequalities to be satisfied by initial data wave function $|\phi^c>$. The higher the order of classicality $L$ the more restrictive is the former system.
For typical values of $\delta_q(0),\delta_p(0)$ (and choosing $L$ of reasonable size) there are many solutions of (61). Gaussian wave packets, for instance, provide well-known solutions:
\begin{equation}
\psi_c(q_0,p_0,\Delta q, q)= \frac{1}{(2\Pi (\Delta q)^2)^{1/4}} \exp \left\{
-\frac{(q-q_0)^2}{4 (\Delta q)^2} + i p_0 q /\hbar \right\}.
\end{equation}
If we take the parameters $q_0$ and $p_0$ to be given by $q_0=q(0)$ and $p_0=p(0)$, 
and substitute $\psi_c(q(0),p(0),\Delta q,q)$ in (61) we get:
\begin{equation}
\frac{(2L-1)!}{2((L-1)!)}(\Delta q)^{2L} \le \delta_q(0)^{2L} \quad \wedge \quad
\frac{(2L-1)!}{2((L-1)!)}\left(\frac{\hbar}{2^{1/2} \Delta q}\right)^{2L} \le \delta_p(0)^{2L}.
\end{equation}
Any Gaussian wave function of the form (62), with the parameter $\Delta q$ satisfying the inequalities (63) for a given $L$, is a $L$-order classical wave function with respect to the classical initial data $\{q(0),p(0),\delta_q(0),\delta_p(0)\}$. Hence, if the classical sector initial time configuration is in such a state then the $L$-order half quantum predictions are valid.  

\section{Conclusions}

The general prescription to derive a theory of coupled classical-quantum dynamics presented in this paper might be summarised in three main steps: 1) Identification of the properties that should be satisfied by the full quantum initial data so that it might be properly described by a set of half quantum initial data (section 2, eq.(3)).
2) Establishment of a relation between a general full quantum observable and the correspondent half quantum one so that one is able to reproduce the predictions of quantum mechanics using the half quantum operators (eq.(10)). 
This evolves the derivation of a relation between the (central) eigenvectores of $\hat B$ and the eigenvectores of $\hat A$ (eq.(19)) and in the sequel of a relation between the probabilities in the representation of $\hat B$ and of $\hat A$ (eq.(39,40)).
3) Finally, the derivation of a framework providing the half quantum operators without requiring previous knowledge of the full quantum theory (section 3).

Certainly, there are many different ways of implementing this general plan (see for instance
\cite{diosi1,nujo}). In this paper we presented a particular derivation of a theory of coupled classical-quantum dynamics that was named half quantum mechanics. This theory, in the form of a set of axioms, was firstly presented in \cite{aleksandrov,boucher}. Its properties have been extensively discussed in the literature \cite{anderson1,diosi1,jones,salcedo1,salcedo2}. In particular, the fact that the bracket structure does not satisfy the Jacobi identity is known to be problematic, the dynamical structure displaying a set of undesirable properties (it is not unitary and time evolution does not preserve the bracket structure, just to mention two of the most intriguing). However, and despite of the fact that the internal structure of half quantum mechanics is not the most desirable, the theory was shown to provide a valid description of coupled classical-quantum dynamics in the sense that it reproduces the results of quantum mechanics in the appropriated limit.   
The key issue in half quantum mechanics is, of course, the way in which its predictions should be interpreted. Associated to every prediction is an error margin, and within this error margin the theory is physically valid. 

To finish we would like to make a few comments: \\
a) There is an uncertainty associated to all  
predictions made by the half quantum theory. Since we do not have a complete 
knowledge of the initial data wave function we could not expect to 
have a complete 
deterministic prediction, much the same to what happens in classical 
mechanics. 
As expected, the degree of precision of the half quantum predictions is related to 
the classicality conditions that are assumed to be satisfied by the classical sector initial data wave function or, in other words, to the degree of validity of the classical initial data.\\
b) A different bracket for classical-quantum dynamics have been presented in the literature \cite{anderson1}. The new theory was also postulated and motivated in terms of its properties. This has caused much debate over which would be the best structure for a theory of coupled classical-quantum dynamics. We would like to point out that Anderson theory might also be obtained through a procedure similar to the one presented in this paper. To do this we just have to use a slightly different unquantization map. The deductive approach will provide a way of comparing the two theories in what respects to their consistency with the full quantum description. \\
c) Lastly, as a side result, we realised that the fact that the brackets do not satisfy the Jacobi identity is clearly a consequence of the fact that the unquantization map is not univocous. This might point out a path to obtain a new, better behaved theory of coupled classical-quantum dynamics \cite{nujo}.

\section*{Appendix - Error ket framework}

The aim of this appendix is just to present some of the
results of the error ket framework. For a more detailed
presentation the reader should refer to \cite{nuno1}.

Let us start by introducing the relevant definitions. Let $\hat{X_i},i=1..n$ 
be a set of $n$ operators acting on the Hilbert space ${\cal H}$ and let $|\psi>$ be the wave function describing the system.

\bigskip 

\underline{{\bf Definition}} {\bf - Error Ket}\\
We define the n-order mixed error ket $|E(\hat{X}_1,\hat{X}_2,...\hat{X}_n,
\psi,x^0_1,x^0_2,...x^0_n)>$, 
as the quantity:
\begin{equation}
|E(\hat{X}_1,\hat{X}_2,...\hat{X}_n,\psi,x^0_1,x^0_2,...x^0_n)> 
=(\hat{X}_1 - x_1^0) (\hat{X}_2 - x_2^0).... (\hat{X}_n - x_n^0) 
|\psi>,
\end{equation}
where $x^0_i$ are complex numbers and the operators $\hat{X}_i$ 
do not need to be self-adjoint.
The error bra $<E(\hat{X}_1...\hat{X}_n,\psi,x_1^0,...x^0_n)|$ 
is defined according to the definition
of the error ket. When there is no risk of confusion we will also use the notation
$|E_{X_1,...X_n}>$ for the mixed error ket. Moreover, when $\hat X_1=\hat X_2 = ... \hat X_n =
\hat X$ the error (64) is named "$n$-order error ket" and we write it in the form: $|E_X^n>=|E^n(\hat X,\psi,x^0)>$. \\ 

Let us present some properties of the former quantity: \\
{\bf a)} The error ket provides a confinement of the wave function. \\
Let $\hat{X}$ be self-adjoint and $x^0 \in {\cal R}$. Given $<E^n_X|E^n_X>$ to each "quantity of
 probability" $p$ we can associate an interval $I$
around $x^0$, $I=[x^0-\Delta_n,x^0+\Delta_n]$, such that the probability 
of obtaining a value $x \in I$
from a measurement of $\hat{X}$ is at least $p$. The 
size of the interval $I$ is dependent of $\hat{X} , \psi$ and $x^0$ only
through the value of $<E^n_X|E^n_X>$. To the quantity $\Delta_n=\Delta_n(\hat{X},\psi,x^0,p)$ we call the $n$-order spread of the wave function. $\Delta_n$ is given by:
\begin{equation}
\Delta_n(\hat{X},\psi,x^0,p)=\left(\frac{<E^n_X|E^n_X>}{1-p}\right)^{1/2n}.
\end{equation}
If $\hat{X}$ is not self-adjoint then 
the former result can also be obtained, but in this case $I$ is a ball of 
radius $\Delta_n$ in the complex plane. \\
{\bf b)} Let $\hat X$ be self-adjoit and $x^0 \in {\cal R}$. The former result can be restated in the following way: given $<E^n_X|E^n_X>$
and a distance $d$, the probability of obtaining a value $x\notin
[x^0-d,x^0+d]$ from a measurement $\hat{X}$ is at the most $<E^n_X|E^n_X>/d^{2n}$. In fact,
(let $|x,k>$ be a complete set of eigenvectores of $\hat X$, where $x$ is the associated eigenvalue and $k$ is the degeneracy index):
\begin{eqnarray}
& & <E^n_X|E^n_X>  =  \sum_{x,k} (x-x^0)^{2n}|<\psi|x,k>|^2 \nonumber\\
& \ge & \sum_{x \notin [x^0-d,x^0+d],k} 
 (x-x^0)^{2n}|<\psi|x,k>|^2 \ge d^{2n} \sum_{x \notin [x^0-d,x^0+d],k} |<\psi|x,k>|^2   ,
\end{eqnarray}
and this implies:
\begin{equation}
\sum_{x \notin [x^0-d,x^0+d],k} |<\psi|x,k>|^2 \le \frac{<E^n_X|E^n_X>}{d^{2n}} = \frac{\Delta_n(p)^{2n} (1-p)}{d^{2n}}.
\end{equation}

\subsection*{Acknowledgements}

I am very grateful to Jorge Pullin for 
encouragement and many discussions 
and suggestions concerning the present paper. 
I would also like to thank Gordon Fleming and Lee Smolin for 
several discussions and insights in the subject.

This work was supported by funds provided by Junta Nacional de 
Investiga\c{c}\~{a}o Cient\'{i}fica e Tecnol\'{o}gica -- Lisbon -- Portugal, 
grant B.D./2691/93 and by grants NSF-PHY 94-06269,
NSF-PHY-93-96246, the Eberly Research fund at Penn State and
the Alfred P. Sloan foundation.


\begin{thebibliography}{99}

\bibitem{anderson1} A. Anderson, {\it Phys. Rev. Lett.} {\bf 74}, 621, (1995).

\bibitem{diosi1} L. Di\'{o}si, N. Gisin, W. T. Strunz, {\it Phys. Rev.} {\bf A61}, 22108 (2000).

\bibitem{aleksandrov} I. V. Aleksandrov, {\it Z. Naturforsch.} {\bf 36A}, 902 (1981).

\bibitem{boucher} W. Boucher, J. Traschen, {\it Phys. Rev.} {\bf D37}, 3522, (1988).

\bibitem{halliwell1} J. J. Halliwell, {\it Phys. Rev.} {\bf D57}, 2337-2348 (1998).

\bibitem{halliwell2} J. J. Halliwell, {\it e-print:} gr-qc/9808071 (1998).

\bibitem{nature} J. Maddox, Nature (London) {\bf 373}, 469 (1995).

\bibitem{sherry} T. N. Sherry, E. Sudarshan, {\it Phys. Rev.} {\bf D18}, 4580-4589 (1978).

\bibitem{diosi2} L. Di\'{o}si, {\it e-print:} quant-ph/9902087 (1999).

\bibitem{hartle1} J.B. Hartle, {\it Spacetime quantum mechanics and the 
quantum mechanics of spacetime} in Gravitation and Quantifications,
eds B. Julia and J. Zin-Justin, Les Houches, Session LVII, (1992).

\bibitem{halliwell3} J. J. Halliwell in {\it Fundamental Problems in Quantum Theory}, edited by D. Greenberger and A. Zeilinger, Annals of the New York Academy of Sciences, Vol. 775, 726, (1994).

\bibitem{wald} R. Wald, {\it Quantum field theory in curve space-times
and black hole thermodynamics}, (Chicago University Press, Chicago, 1994)
and references therein.

\bibitem{nuno1} N.C. Dias, {\it e-print:} quant-ph/9912034 (1999).

\bibitem{dirac1} P.A.M. Dirac, {\it The principles of Quantum Mechanics},
(Clarendom Press, Oxford, 1930).

\bibitem{dirac2} P.A.M. Dirac, {\it Lectures on Quantum Mechanics},
Yeshiva University, (Academic Press, New York, 1967).

\bibitem{nujo} N. Dias, J. Prata, {\it e-print:} quant-ph/0005019 (2000).

\bibitem{jones} K. R. W. Jones, {\it Phys. Rev. Lett} {\bf 76}, 4087 (1996); L. Di\'{o}si, ibid. p4088; I. R. Senitzky ibid. p4089; A. Anderson ibib. p4089-4090.

\bibitem{salcedo1} L. L. Salcedo, {\it Phys. Rev.} {\bf A54}, 3657 (1996).

\bibitem{salcedo2} J. Caro, L. L. Salcedo, {\it Phys. Rev.} {\bf A60}, 842 (1999).







\end{thebibliography}
\end{document}